# Comprehensive Analysis for the High Field Magneto-conductivity of $Bi_2Te_3$ Single Crystal

Yogesh Kumar[1,2], Rabia Sultana[3] and V.P.S. Awana[1,2*]

[1]*CSIR-National Physical Laboratory, Dr. K.S. Krishnan Marg, New Delhi-110012, India*
[2]*AcSIR-National Physical Laboratory, New Delhi-110012, India*
[3]*Asutosh College, University of Calcutta, Kolkata, India-700026*

**Abstract**

Here, we report the magneto-conductivity (up to 14Tesla and down to 5K) analysis of $Bi_2Te_3$ single-crystal. A sharp magneto-conductivity (MC) rise (inverted v-type cusp) is observed near H=0 due to the weak antilocalization (WAL) effect, while a linear curve is observed at higher fields. We account for magneto-conductivity (MC) over the entire range of applied magnetic fields of up to 14Tesla and temperatures from 100K to 5K in a modified HLN modelling (addition of quadratic ($\beta H^2$) through quantum and classical components involvement. The additional term $\beta H^2$ reveals a gradual change of a (HLN parameter) from -0.421(6) to -0.216(1) as the temperature increases from 5 to 100K. The phase coherence length $L_\varphi$ obtained from both conventional and modified modelling decreased with increasing temperature but remains more protracted than the mean free path (L) of electrons. It shows the quantum phase coherence effect dominates at high temperatures.

**\*Corresponding Author**
Dr. V. P. S. Awana,
National Physical Laboratory (CSIR)
Dr. K. S. Krishnan Road, New Delhi-110012, India
E-mail: awana@nplindia.org
Ph. +91-11-45609357, Fax-+91-11-45609310
Home page: awanavps.webs.com

**Introduction**

Topological Insulators (TIs) are the new class of materials, which shows the insulator type band gap in bulk and conductive nature at the surface due to the surface or edge state (SS). These states occur due to the combination of strong spin-orbit coupling (SOC) and time-reversal symmetry (TRS) [1,2]. The presence of strong SOC causes the band inversion due to which gapless SS is formed [3-5].TIs show the odd number of left or right moving edge states, which correspond to an odd number of Dirac cones and an adiabatically moving electron around the Fermi surface. These electrons acquired a $\pi$ Berry phase [6-9]. Due to the presence of SS, TIs show peculiar properties as compared to metals and insulators. Hence TIs are attractive for potential applications. The $Bi_2Te_3$ has been confirmed to possess these topological properties from the first principle study [10], which is further substantiated through experimental observation of the surface Dirac dispersion using angle-resolved photoelectron spectroscopy (ARPES) [11]. The existence of Shubnikov-de Hass (SdH) oscillations due to the topological surface states in $Bi_2Te_3$ crystals has also been revealed, which appears when the energy spacing between two successive Landau level exceeds their broadening due to disorder [12, 13]. Recently a magneto-transport study has also been a prominent way to probe these surface states in TIs [14-18]. The variation in MC with field leads to interesting observations, such as non-saturating linear trends at higher fields and quadratic at lower fields [16, 19-20].

In addition to linear MC at higher fields in bismuth-based layered compounds, the superconductivity has also been observed by doping at low temperatures [21]. Interestingly, unlike the low-field WAL, this novel linear-like MC is less studied. Concerning magneto-conductivity (MC), the nature of scattering governs the type of conduction mechanism in the system as the moving electrons in materials undergo various scattering through electron-electron, electron-phonon scattering, or defects scatterings. Competition between the mean free path (L) and phase coherence length ($L_\varphi$) modifies the channel of conductivity. If $L_\varphi \leq L$, then electrons are in a diffusive regime, and they follow the Drude conductivity. While if $L_\varphi >> L$, then electrons are in a quantum diffusive regime, and correction in conductivity leads to weak localization (WL) or weak anti-localization (WAL) effects [12]. The $L_\varphi$ is temperature-dependent, and it increases with a decrement in temperature. Also, the applied magnetic field suppresses the quantum effects too; hence WL and WAL effects are observed at low temperatures and fields [15-18]. The WL and WAL give positive and negative contributions to magneto-conductivity at low magnetic fields and temperatures, respectively [22-25]. WAL decreases the conductivity at low temperatures as backscattering is diminished by the π Berry phase along the Fermi surface [16, 26- 27]. The MC can be described through the Hikami-Larkin-Nagaoka (HLN) equation. However, the HLN fitted curve diverged from experimentally observed magneto-conductivity at a higher temperature and applied magnetic field as spin-orbit scattering time becomes smaller than elastic scattering time [28]. The HLN model fitting effectively studied the signature of WAL in 2D systems of the magneto-conductivity at low temperature and applied magnetic fields [20]. The HLN parameters viz., pre-factor (α) and $L_\varphi$ gives us information about the type of localization and number of conducting carrier channels [29]. As the pre-factor (α) is negative for WAL and positive for WL. At lower magnetic field and temperature, the inelastic scattering term will be dominant, as $L_\varphi$ is more generous than L [30]. Previously HLN analysis was done to explain the surface conductivity of various TIs. However, it was mainly for TI thin films and nanostructures [31, 32], and the analysis was limited by their thickness and size-dependent quantum confinement effects, being added with intrinsic surface conductivity of a TI. Apart from these low field MR, an intense linear-like MR at a high magnetic field of 60 Tesla is reported in the Bi2Te3 thin films, which are deposited through the pulsed laser deposition (PLD) technique [28]. It suggests the presence of enormous MR due to WAL phenomena as in bulk single crystals [15-18, 28]. The theoretical analysis of the PLD deposited films confirms the longer time of elastic scattering than the spin-orbit scattering time [28]. Although there are reports related to HLN being applied to bulk single crystals of TI, but in all these cases mostly are in a low field, and temperature regime (≈ 1Tesla, 2-10K) [33-36], and its applicability with appropriate quantum modifications and bulk contribution to overall conductivity in higher fields and temperatures regime is scant.

Earlier, we have reported HLN analysis of magneto-conductivity of bulk $Bi_2Te_3$ single crystal [37] up to ±5Tesla field range. In the current article, we report both conventional HLN and modified HLN treatment up to ±14Tesla field by adding a quadratic term ($\beta H^2$), accounting for elevated temperatures and higher magnetic fields. The sign of the coefficients of these terms determines the type of quantum and classical corrections needed in HLN magneto-conductivity [22, 38]. Here, we show that the quantum phase coherence effect dominates at the higher temperatures (up to 50K). It indicates that the electron retains their quantum phase coherence ($L_\varphi$) up to a distance, which is more significant than the mean free path of the electron (L). At even higher temperatures (100K) and higher fields (>10Tesla), the bulk carriers also contribute to conductivity. The classical correction term (γH) and quadratic term ($\beta H^2$) are added to the conventional HLN model to account for temperatures (5K to 100K) and fields (up to 14Tesla) in order to explain the magneto-conductivity of a bulk single crystalline $Bi_2Te_3$ topological insulator.

**Experimental Details**

Single crystal of bismuth telluride ($Bi_2Te_3$) was grown using solid-state reaction by self-flux method via vacuum encapsulation process in automated programmable furnace. High purity (99.99%) bismuth (Bi) and tellurium (Te) powders were taken in the appropriate stoichiometry and ground thoroughly using mortar & pestle in inert argon (Ar) atmosphere glove box to avoid any contamination or oxidation from the surrounding environment. The homogeneously mixed powder of Bi and Te were pressed into a rectangular pellet form by applying hydraulic pressure (100kg/cm$^2$) with the help of a pelletizer, followed by the vacuum encapsulation (<10$^{-5}$mbar) of the pellet in a quartz tube. The encapsulated quartz tube containing the rectangular pellet was kept in an automatic programmable electric box furnace under an optimized heat treatment as follows, the sample was heated up to 950°C with a heating rate of 120°C/hour [37, 39]. With the aim of getting a good homogeneity of the mixture in the molten state, the furnace was allowed to remain at the same temperature (950°C) for 12 hours. Subsequently, the furnace was cooled down very slowly from 950°C to 650°C with a cooling rate of 2°C/hour so that the atoms could occupy their respective positions. Finally, the furnace was naturally cooled to room temperature [37]. The resultant sample was taken out by breaking the encapsulated quartz tube, and mechanically cleaved flakes of obtained single crystal were used for further measurements. X-ray diffraction (XRD) was performed on Rigaku made MiniFlex - II with Cu-K$_\alpha$ radiation having a wavelength ($\lambda$) of 1.5418 Å. The magneto-transport measurements were performed on Quantum Design Physical Property Measurement System (PPMS-14Tesla down to 2K) using the standard four-probe method.

**Results and Discussion**

The surface X-ray diffraction spectra of Bi$_2$Te$_3$ single-crystal flakes (shown in Figure 1) are recorded for 2θ ranges 10° to 80° in atmospheric conditions. From the recorded data, the single crystalline property is confirmed and it is observed that the growth of the crystal is along the (00$l$) diffraction plane, having a rhombohedral crystal structure of $R\bar{3}m$ space group (for details, see ref. 39, reported earlier by some of us). The behaviour of electrical resistivity (ρ) of Bi$_2$Te$_3$ as a function of temperature (ranging from 50K to 5K) under different applied magnetic fields (up to 10Tesla) is shown in Figure 2(a), which is measured using the standard four-probe method (schematically shown in the inset of Fig. 2(a)). At a particular magnetic field, the resistivity of Bi$_2$Te$_3$ single crystal decreases with decreasing temperature, exhibiting both metallic as well as positive transverse MR (magneto-resistance) behaviour. This positive MR originates due to the change of significant electronic trajectories by the action of the magnetic field. The MR is usually calculated as the ratio of resistivity with the change in the magnetic field;

$$MR\ (\%) = \frac{\rho(H) - \rho(0)}{\rho(0)} \times 100 \qquad (1)$$

Figure 2(b) shows the variation of MR(%) of Bi$_2$Te$_3$ with the applied magnetic field (H) up to 14 Tesla at different temperatures. At the lowest studied temperature (5K), a linear curve is observed from the quantum magneto-conductivity and classical contribution combinations. Further, as the temperature increases from 5 to 100K, the classical magneto-conductivity dominates, resulting in the parabolic appearance, i.e., a broad cusp. The magnetic field dependence of MR showed two distinct behaviours, a quadratic term at the lower field and a non-saturating linear-like trend at the higher fields [38, 40]. The MR vs H plot indicates a large linear magneto-resistance (LMR), i.e., 434%, 14Tesla at a low-temperature of 5K, which decreases to 145% at 100K. This phenomenon was explained in detail by the Abrikosov model, which accounted for linear MR even in the lower fields. Sometimes the disorder induced by the in-homogeneity in materials leads to a zero-gap state. This gapless dispersion is due to the extreme quantum limit, where all electrons coalesce into the lowest Landau level [41].

Further, in order to explain the MR behaviour classically, a phenomenological model by Parish and Littlewood (PL) was set forth [42-44]. It explained a proportionality relationship between carrier mobility, sample in-homogeneity, and LMR. This purely classical geometrical effect distresses the fluctuations in carrier mobility and thus attributes to the linear non-saturating MR at the higher fields. However, the PL model was restricted to disordered systems only. The presently studied Bi$_2$Te$_3$ single crystals is a well-characterized [39, 45] material, with an ordered layered structure as evidenced by scanning electron microscopy. Further, Bi$_2$Te$_3$ is a proven intrinsic quantum material [1-10]. Therefore, the PL model may not be considered as the prime reason for such a huge LMR. Henceforth, the HLN model is considered to probe the observed MR, which is further modified to incorporate various other contributions.

According to HLN, in the presence of the magnetic field, the change in conductivity of a 2D electron system exhibiting strong SOC at different temperatures (5, 50, and 100K) and magnetic field (up to ±14Tesla) as shown in Figure 3 (a) is given by:

$$\Delta\sigma(H) = -\frac{\alpha e^2}{\pi h}\left[ln\left(\frac{B_\varphi}{H}\right) - \Psi\left(\frac{1}{2} + \frac{B_\varphi}{H}\right)\right] \quad (2)$$

where e is the electronic charge, h is Plank's constant, H is applied magnetic field, $\Psi$ is digamma function, $B_\varphi = \frac{h}{8e\pi L_\varphi^2}$ is characteristic field, and L$_\varphi$ is phase coherence length. Here the magneto-conductivity data is calculated from the inversion of resistivity data. The difference of magneto-conductivity ($\Delta\sigma$) is given by $\Delta\sigma(H) = \sigma(H) - \sigma(0)$. Where [$\sigma$(H)] is the magneto-conductivity at the applied magnetic field and [$\sigma(0)$] at zero fields.

The HLN parameters $\alpha$ and L$_\varphi$ are used to determine the type of scattering or localization and number of 2D coherent conducting channels present in the material. In the HLN model, depending upon spin-orbit interaction (SOI) and magnetic scattering, the value of $\alpha$ varies as -1/2 (symplectic case), 0 (unitary case), 1 (orthogonal case) [19, 20] and L$_\varphi$ is determined by inelastic scattering from electron-electron interaction or the electron-phonon coupling. From Fig. 3(a), it is clear that HLN model fitted curves resemble the experimental data at lower applied magnetic fields but diverges at higher magnetic fields. It limits the HLN model to a lower magnetic field range. The values of fitted parameters $\alpha$ and L$_\varphi$ for the full range of applied magnetic field (±14Tesla) at different temperatures are given in Table 1. The value $\alpha$ lies within the range of -0.335(9) to -0.285(1) for temperature range 5-100K, signifying single surface state transport with a major bulk contribution to the conductivity. Apparently, with increasing temperature from 5K to 100K, there is no effective change in the value of $\alpha$ but phase coherence length (L$_\varphi$) changes from 49.647(2) nm to 16.222(2) nm. The earlier report indicated a similar trend with $\alpha$ nearly unchanged and remaining within the range of from -0.855 to -0.88 and L$_\varphi$ from 96.045 nm to 40.314 nm with temperature increasing from 5 to 50K [37]. Here the magneto-conductivity data was HLN fitted at much lower magnetic fields, i.e., ±0.25Tesla [37]. However, as comparing the HLN fitting of the PLD thin films, it is evident that these films show a strong WAL effect as the value of $\alpha$ (1.5 K) is -0.711 and $\alpha$ (30 K) is -0.88 [28]. While the present single crystal shows a weak WAL effect as shown by the values in table 1. These parameters also confirm a mechanism close to the single conducting channel in synthesized crystals, but the thin films show a close contrivance to the double channel [28]. We can say that there are two surfaces conducting channel (for the low field; ±0.25Tesla) and a single surface conducting channel (for the high field; ±14Tesla) along with a major bulk component (WL) which contributes to the conduction mechanism. As the strength of the applied field is increased, one cannot ignore the classical contribution to the conductivity. Hence, the HLN equation has to be modified for a complete analysis of magneto-conductivity (MC) in higher magnetic field ranges.

The electron undergoes elastic scattering from static scattering centers at a higher field, which can be approximated by H$^2$. Hence the $\beta$H$^2$ term is added in the conventional HLN equation [14, 46]. Here, $\beta$ represents both quantum and classical contribution ($\beta_q + \beta_c$) to the magneto-conductivity. It includes the contribution from characteristic magnetic fields corresponding to spin-orbit scattering length (L$_{so}$) and elastic scattering length (L$_e$). A schematic representing elastic and inelastic scattering at the low and high magnetic field is shown in Figure 3(b). At the small magnetic field, the coherent scattering term dominates as the $L_\varphi$ is larger than $L_{so}$ and $L_e$. While at a higher field regime, $L_{so}$ and $L_e$ become essential, as the elastic scattering and cyclotronic scattering term also contribute to MC and HLN. Hence, the resulting MC equation is given as:

$$\Delta\sigma(H) = -\frac{\alpha e^2}{\pi h}\left[\ln\left(\frac{B_\varphi}{H}\right) - \Psi\left(\frac{1}{2} + \frac{B_\varphi}{H}\right)\right] + \beta H^2 \qquad (3)$$

Where the first term $\left\{-\frac{\alpha e^2}{\pi h}\left[\ln\left(\frac{B_\varphi}{H}\right) - \Psi\left(\frac{1}{2} + \frac{B_\varphi}{H}\right)\right]\right\}$ belongs to the conventional HLN equation, and the parameters are defined above, while β is a coefficient of the quadratic term which arises from elastic scattering term. The fitting parameters are α, β, and $L_\varphi$ and values of these parameters are given in Table 2, along with the R-square value indicating the goodness of fitting. At 5K, the pre-factor α is obtained to be -0.421(6), which slightly differs from -0.5 (symplectic case), showing negligibly weak magnetic scattering, and SOI is strong. In other words, we can say that at 5K, the conductivity mechanism exhibits a single surface state dominated transport along with minor bulk contribution and thus, confirms the 2D nature of WAL. Further, the value α is observed to gradually change from -0.421(6) to -0.216(1) as the temperature increases from 5 to 100K. The significant increase in the value α is possibly due to the additional bulk channel partial coupling with the surface conducting channel. The increase in α value with increasing temperature clearly reveals that the surface contribution to the conductivity mechanism decreases with increasing temperature. It also suggests that a major bulk effect comes into play as the temperature increases from 5 to 100K. On the other hand, $L_\varphi$ decreases from 43.934(5) nm to 17.518(2) nm as the temperature increases from 5K to 100K. The coefficient of the quadratic term β has minimal value (~ $10^{-4}$) and, as predicted, β comes out to be positive because a positive correction needs to be added in HLN, which lies below the experimental data. At 5K, although the HLN fitted curve is observed to be above the experimental data, β still appears to be positive. It is because when the modified HLN fitting is carried out, α becomes -0.421(6) which further, drags the HLN curve below the experimental data. Fig.3 (a) shows that by adding the $\beta H^2$ term in the conventional HLN model equation, the fitted curve completely resembles the experimental data at all temperatures. The most interesting part is that the above-modified HLN equation requires dominancy of the quantum coherence effect [38], indicating that electrons retain their quantum phase coherence even at such high temperatures (100K), i.e., $L_\varphi$>>L. Hence here we account for LMR dominancy in the quantum diffusive regime.

Till now, it is clear that for the lower magnetic field (< 1Tesla) and temperature, the conventional HLN equation explains the WAL behaviour, but deviates from experimental data at a higher magnetic field. The addition of the $\beta H^2$ term in the conventional HLN equation describes the quantum scattering and classical cyclotronic contribution. As the temperature is increased, the bulk contribution to the conductivity of the material comes into the picture [47, 48]. In TIs, based on the strength of the magnetic field and temperature, the conductivity is governed by both surface and bulk charge carriers. Surface charge carriers dominate at the lower field and lower temperatures, easily understood through figure 4. As the field strength and temperature are increased, the bulk carriers start contributing to conductivity. A recent report [49] suggested the signature of bulk contribution in TI conductivity at extreme conditions viz. magnetic field, pressure, or temperature. There is a competition between surface and bulk carriers. To probe this individually, all the components, i.e., HLN (surface conductivity), $\beta H^2$ (quantum and classical contribution), and a new term γH (classical transport contribution) is added upon to account for all fields and temperature magneto conductivity (MC) of an intrinsic bulk TI. Fig. 4 shows the theoretical fitting of the experimental data with these proposed terms. First, we observed that HLN fitted curve deviates at fields around 1.5 Tesla from experimental data, which is in accordance with previous reports. Second, when alone the quadratic term $\beta H^2$ is used to fit the data, it is observed that it deviates from experimental data at the lower magnetic field ($\leq$ 3 Tesla) and also at higher fields above 10 Tesla, which is due to the fact that quantum diffusive effects are dominating in mid fields (but above HLN regime) and classical effect starts to be effective at higher fields. Lastly, the data is fitted with $\gamma H$ term, which is well fitted at higher fields ($\geq$ 10 Tesla), indicating bulk classical contribution to overall magneto-conductivity. Hence it is clear from Figure 4 that various field regions are dominated by the contributions of different terms, i.e., HLN, quadratic ($\beta H^2$), and classical ($\gamma H$). Here, to study the bulk contribution to the surface conductivity, a term proportional to the applied field is further added to HLN.

$$\Delta\sigma(H) = -\frac{\alpha e^2}{\pi h}\left[\ln\left(\frac{B_\varphi}{H}\right) - \Psi\left(\frac{1}{2} + \frac{B_\varphi}{H}\right)\right] + \beta H^2 + \gamma H \qquad (4)$$

Here the contribution of the first two terms is already explained, while the third term indicates the bulk contribution in surface conductivity. The value of $\gamma$ governs the strength of bulk contribution. The above equation is fitted for the entire field range (up to 14Tesla) at temperatures 5, 50, and 100K, see Figure5. At 5K, the value of $\alpha$ drastically changes and becomes positive by the addition of the $\gamma H$ term (see Table 3). Interestingly, the positive value of $\alpha$ is not acceptable within HLN. It calls for the non-applicability of the $\gamma H$ term at 5K. Bulk carriers do not contribute to the overall conductivity of $Bi_2Te_3$ at 5K. At 50K, the value of $\alpha$ changes to -0.409(7), indicating two surface channel contribution along with the quadratic ($\beta H^2$) and linear ($\gamma H$) contributions. At further higher temperatures (100K), the value of $\alpha$ increases from -0.409(7) to -0.215(1), which is reasonable within HLN, as the number of surface channels decreases with an increase in temperature. The role of bulk carriers in surface conductivity increases with increment in temperature. It is clear from Table 1, 2, and 3 that at different temperatures and applied fields, the HLN, quantum scattering, classical ($\beta H^2$), and classical ($\gamma H$) contributions do compete with each other.

**Summary and Conclusion**

In conclusion, we show the magneto-conductivity analysis of $Bi_2Te_3$ single crystal at various temperatures from 5 to 100K and under applied magnetic fields of up to 14Tesla. It is found that HLN alone cannot account for the studied temperature (5 to 100K) and field (up to 14Tesla) magneto-conductivity of $Bi_2Te_3$ but rather, the addition of quadratic ($\beta H^2$)contributions are needed. Further, the HLN, quantum scattering, and cyclotronic motion ($\beta H^2$), and classical ($\gamma H$) contributions compete with each other at 5K as the analysis with the addition of the third term (linear in H) appears unphysical for higher temperatures. While the former is dominant at low T (5K) and fields (<1Tesla), the later start playing their roles at relatively higher temperatures (50, 100K) and fields (>2Tesla).

**Acknowledgment**

The authors would like to thank the Director of National Physical Laboratory (NPL), India, for his keen interest in the present work and CSIR, India, for a research fellowship, and AcSIR-NPL for Ph.D. registration. The author would like to acknowledge Mr. Prince Sharma for his contribution to manuscript writing.

**Figure Captions**

Figure 1:X-ray diffraction spectra of $Bi_2Te_3$ single crystal and the crystal growth is along (00$l$) diffraction plane.

Figure 2(a): Electrical resistivity versus temperature at the different applied magnetic fields. The inset shows the schematic diagram for the magneto-transport measurement by using the four-probe method.

Figure 2(b): MR (%) versus the applied magnetic field at different temperatures.

Figure 3(a): Magneto-conductivity data for $Bi_2Te_3$ single crystal as a function of the magnetic field at different temperatures fitted using the conventional and modified HLN equation by the addition of $\beta H^2$ term.

Figure 3(b): Conduction mechanism as a function of the applied magnetic field. (a) At zero applied magnetic fields, spin-orbit interaction dominates. (b) As the strength of the applied magnetic field is increased, the electrons undergo elastic scattering also.

Figure 4: Magneto-conductivity data depicting the competing contributions of different terms as a temperature and magnetic field function.

Figure 5: Magneto-conductivity of $Bi_2Te_3$ single crystal is fitted using both conventional and modified HLN equation by the addition of $\beta H^2$ and $\gamma H$ term.

**Table Captions**
Table 1: HLN fitting parameters of $Bi_2Te_3$ single crystal at a low magnetic field.
Table 2: HLN + $\beta H^2$ fitting parameters of $Bi_2Te_3$ single crystal up to 14Tesla
Table 3: HLN + $\beta H^2$ + $\gamma H$ fitting parameters of $Bi_2Te_3$ single crystal up to 14Tesla

Table 1

| T (K) | α | $L_\varphi$ (nm) | R-square |
|---|---|---|---|
| 5 | -0.335(9) | 49.647(2) | 0.9976 |
| 50 | -0.376(8) | 24.426(2) | 0.9994 |
| 100 | -0.285(1) | 16.222(2) | 0.9985 |

Table 2

| T (K) | α | $L_\varphi$ (nm) | β | R-square |
|---|---|---|---|---|
| 5 | -0.421(6) | 43.934(5) | 2.307 × $10^{-4}$ | 0.9997 |
| 50 | -0.339(4) | 25.779(7) | 1.446 × $10^{-4}$ | 0.9999 |
| 100 | -0.216(1) | 17.518(2) | 4.690 × $10^{-5}$ | 1 |

Table 3

| T (K) | α | $L_\varphi$ (nm) | β | γ | R-square |
|---|---|---|---|---|---|
| 5 | 5.806(1) | 9.472(3) | 2.099 × $10^{-3}$ | -0.135(8) | 0.9999 |
| 50 | -0.409(7) | 24.928(4) | -9.411 × $10^{-6}$ | 4.957 × $10^{-3}$ | 1 |
| 100 | -0.215(1) | 17.516(6) | 4.831 × $10^{-5}$ | -5.154 × $10^{-5}$ | 0.9999 |

Figure 1

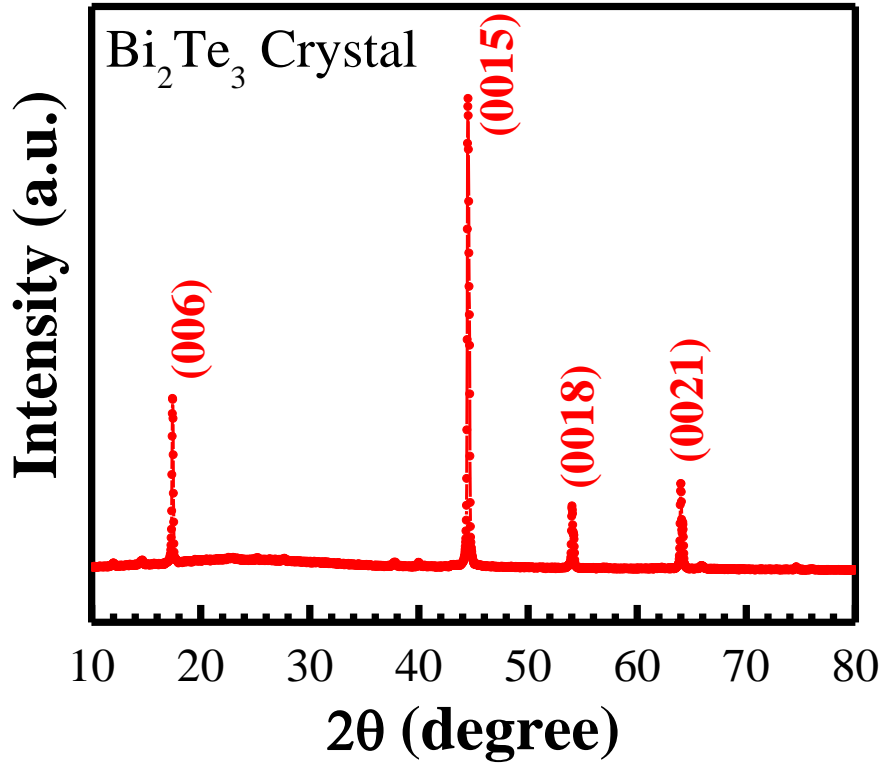

Figure 2(a)

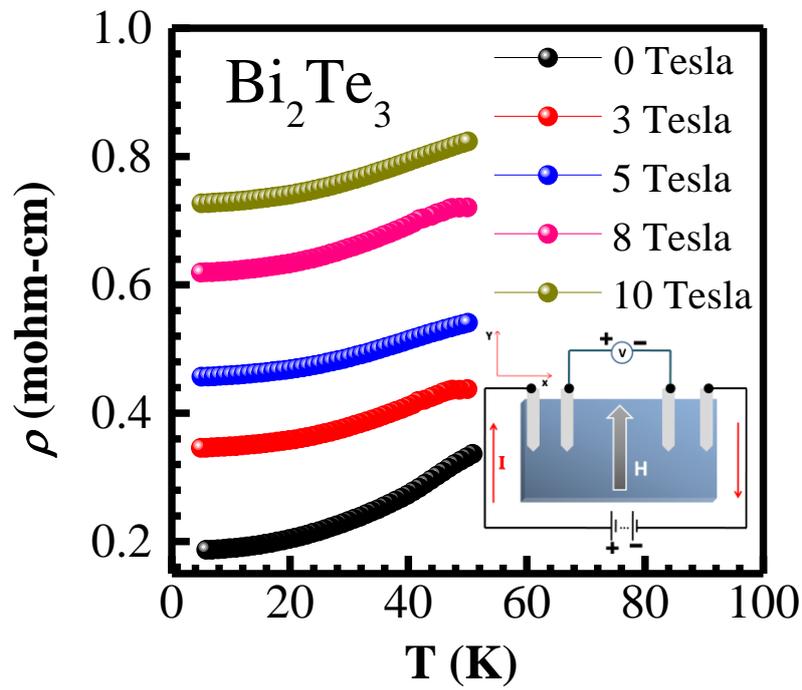

Figure 2(b)

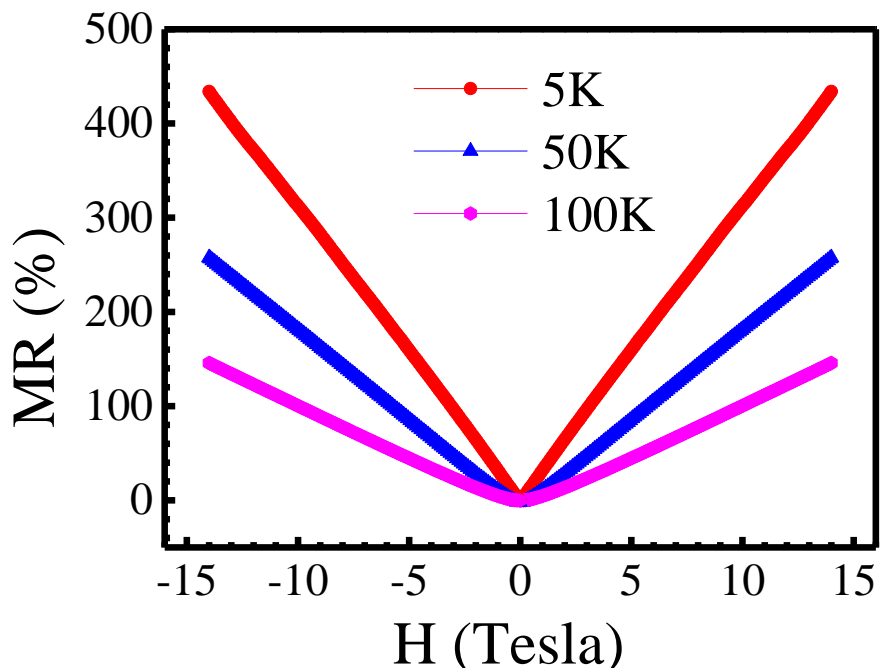

Figure 3(a)

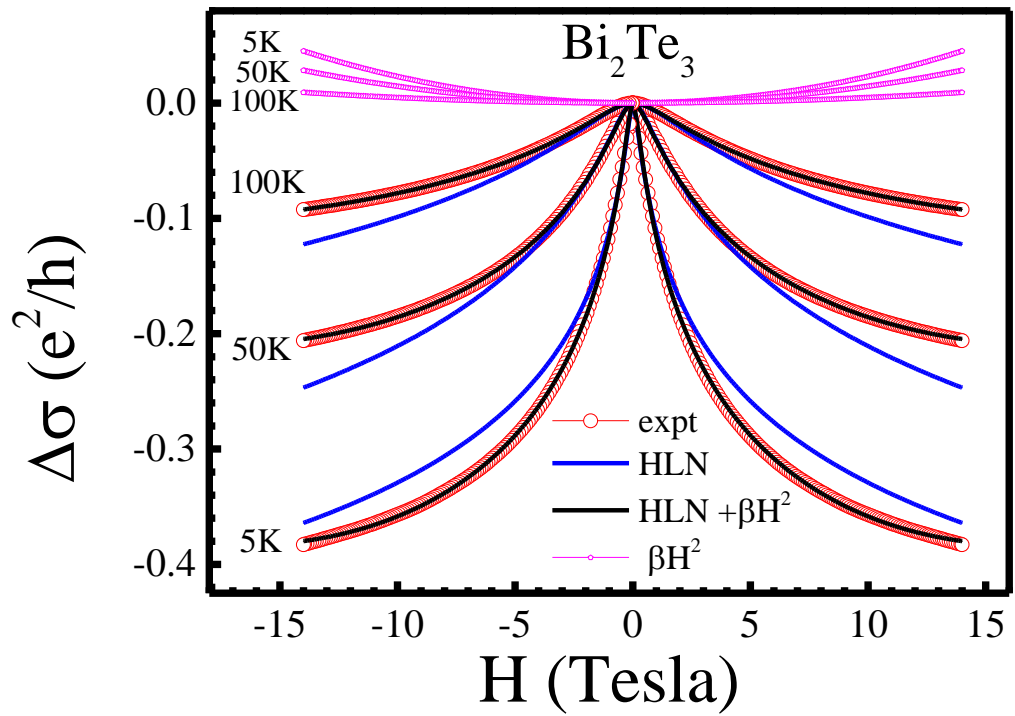

Figure 3(b)

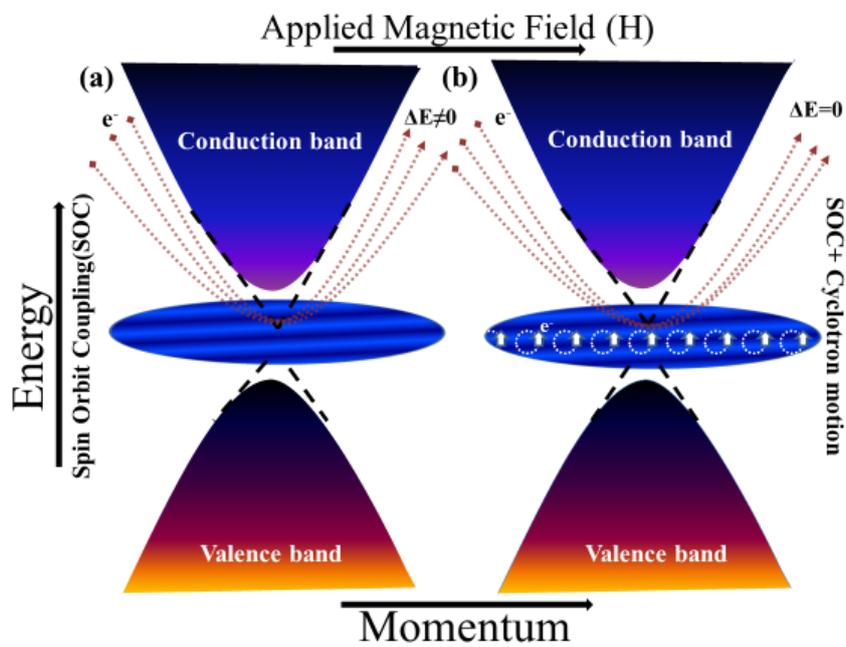

Figure 4

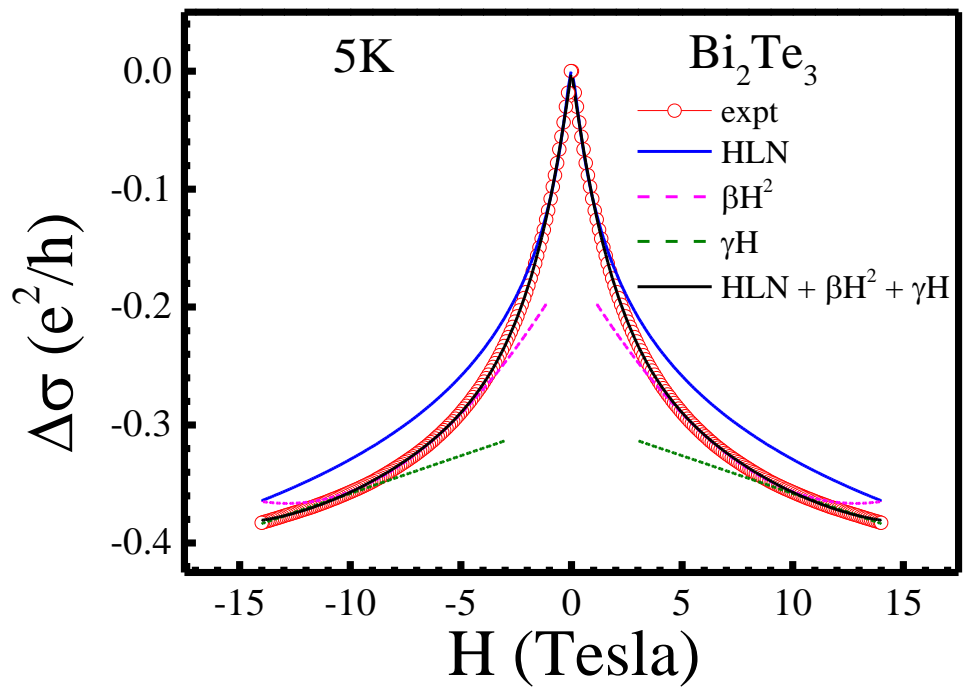

Figure 5

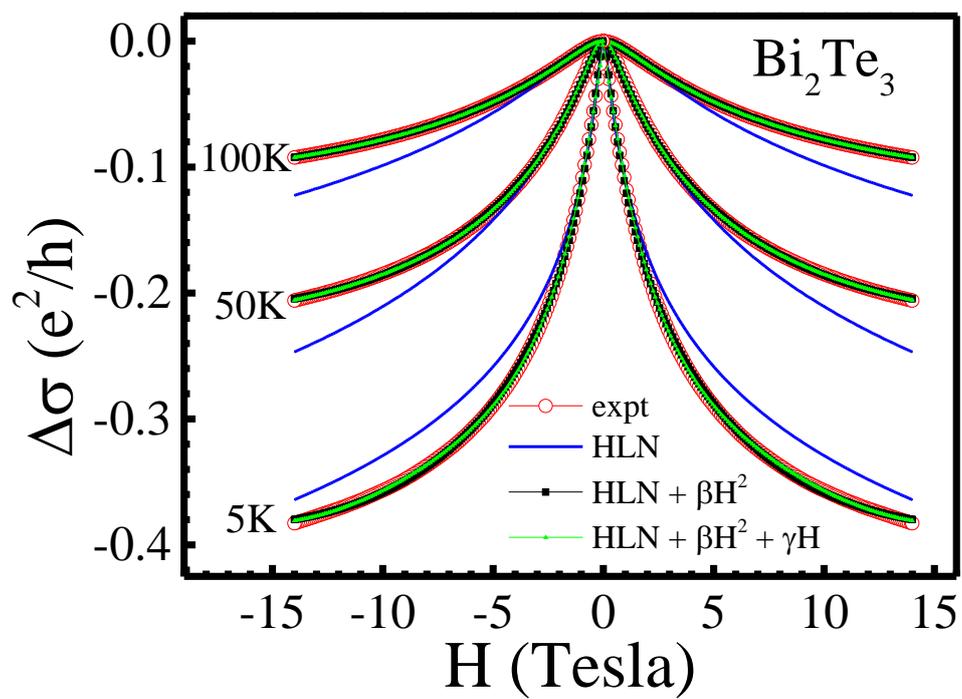